\documentclass[letterpaper,english,aps,prb,floatfix,twocolumn,amsfonts,superscriptaddress,longbibliography]{revtex4-2}

\usepackage[T1]{fontenc}
\usepackage[applemac]{inputenc}
\usepackage[english]{babel}
\usepackage{graphics}
\usepackage{amssymb}
\usepackage{amsmath}
\usepackage{overpic}
\usepackage{natbib}

\usepackage{xcolor}
\usepackage{soul}

\newcommand{\be}{\begin{equation}}
\newcommand{\ee}{\end{equation}}
\newcommand{\bea}{\begin{eqnarray}}
\newcommand{\eea}{\end{eqnarray}}

\usepackage{dcolumn}
\usepackage{bm}
\usepackage{color}
\usepackage{float}

\def\a{\alpha}
\def\b{\beta}
\def\e{\varepsilon}
\def\d{\delta}
\def\g{\gamma}

\def\o{\omega}

\def\s{\sigma}

\def\O{\Omega}

\def\pll{\parallel}

\def\pd{\partial}

\def\bk{{\bf k}}

\def\bq{{\bf q}}

\def\bA{{\bf A}}

\def\nn{\nonumber}
\def\lb{\label}
\def\pref#1{(\ref{#1})}

\newcount\bozza \bozza=0
\ifnum\bozza=1
\newdimen\shift \shift=-2truecm
\def\lb#1{%
{\label{#1}\rlap{\kern\shift{$\scriptstyle#1$}}}}
\else\def\lb#1{\label{#1}} \fi

\begin{document}
\title{THz non-linear optical response in cuprates:\\ predominance of the BCS response over the Higgs mode } 
\author{M.~Udina}
\email{mattia.udina@uniroma1.it}
\affiliation{Department of Physics and ISC-CNR, ``Sapienza'' University of Rome, P.le Aldo Moro 5, 00185, Rome, Italy}
\author{J.~Fiore}
\affiliation{Department of Physics and ISC-CNR, ``Sapienza'' University of Rome, P.le Aldo Moro 5, 00185, Rome, Italy}
\author{T.~Cea}
\affiliation{Imdea Nanoscience, Faraday 9, 28015 Madrid, Spain}
\author{C.~Castellani}
\affiliation{Department of Physics and ISC-CNR, ``Sapienza'' University of Rome, P.le Aldo Moro 5, 00185, Rome, Italy}
\author{G.~Seibold} 
\affiliation{Institut f\"ur Physik, BTU Cottbus-Senftenberg, PBox 101344, 03013 Cottbus, Germany}
\author{L.~Benfatto} 
\email{lara.benfatto@roma1.infn.it}
\affiliation{Department of Physics and ISC-CNR, ``Sapienza'' University of Rome, P.le Aldo Moro 5, 00185, Rome, Italy}
 

\begin{abstract}
Recent experiments with strong THz fields in unconventional cuprates superconductors have clearly evidenced an increase of the non-linear optical response below the superconducting critical temperature $T_c$. As in the case of conventional superconductors, a theoretical estimate of the various effects contributing to the non-linear response is needed in order to interpret the experimental findings. Here we report a detailed quantitative analysis of the non-linear THz optical kernel in cuprates within a realistic model, accounting for the band structure and disorder level appropriate for these systems.  We show that the BCS quasiparticle response is the dominant contribution for cuprates, and its polarization dependence accounts very well for the third-harmonic generation measurements. On the other hand, the polarization dependence of the THz Kerr effect is only partly captured by our calculations, suggesting the presence of additional effects when the system is probed using light pulses with different central frequencies. 
\end{abstract}

\maketitle

\section{Introduction}

The recent technological advances in the generation of strong THz pulses triggered an intense activity aimed at using light to selectively excite fundamental modes in condensed-matter systems\cite{nelson_review13,zhang_review17,cavalleri_review}. In particular, the THz range is the relevant frequency window for phononic excitations and collective modes in broken-symmetry states, like e.g.\ magnons in the magnetically-ordered phase and complex (amplitude and phase) fluctuations of the superconducting (SC) order parameter. In the case of phonons, it has been rapidly understood that experiments with THz pulses\cite{kampfrath_prl17,johnson_prl19} closely mirror the experiments done with intense near-infrared (NIR) and visible (VIS) light fields\cite{giannetti_review}. This is e.g.\ the case for pump-probe protocols, where a weak probe pulse with a variable time delay $t_{pp}$ with respect to the pump detects the relative changes in the reflectivity or transmission through the sample, which can be periodically modulated due to the excitation of Raman-active phonons. In the case of eV light, the interaction between ultrashort optical pulses and lattice degrees of freedom in non-absorbing materials has been unambiguously interpreted\cite{giannetti_review, merlin_ssc97} as an impulsive-stimulated Raman scattering (ISRS) process, where the phonon generation occurs at the difference frequency of two high-energy photons taken from the pump field. In close analogy, experiments done with THz pulses achieve the same goal by a sum-frequency process\cite{kampfrath_prl17,johnson_prl19}, leading to a sharp response whenever the pump frequency $\Omega$ matches half of the phonon energy $\omega_{ph}$, i.e.\ $\Omega=\omega_{ph}/2$. 

The case of Raman-active phonons provides a benchmark example of a more general mechanism allowing for the impulsive excitation of many different modes with a Raman-like symmetry. Indeed, 
as we will discuss in details in this manuscript, the experimental findings can be understood in a rather general way by considering that a Raman-active phonon leads to a strong resonance in the non-linear optical kernel $K(\omega)$ at the phonon frequency $\omega_{ph}$. 
As a consequence, one would expect the same reasoning to hold also for e.g.\ SC collective excitations\cite{cea_prb16,udina_prb19}, once that $\omega_{ph}$ is replaced by the characteristic energy scale $\omega_{res}$ where the corresponding non-linear kernel is resonant. However, even within this scheme the interpretation of experiments in this class of materials has been much more controversial\cite{shimano_review19}. The reason is that even though all measurements in conventional superconductors\cite{shimano_prl12,shimano_prl13,shimano_science14,shimano_prb17,giorgianni_natphys19,wang_natphot2019,wang_mgb2_prb21} can be reconciled with a non-linear kernel peaked at $\omega_{res}=2\Delta$, where $\Delta$ is the SC gap, the identification of the relevant excitations responsible for such resonance has been debated. Indeed in a superconductor both the BCS response probing the quasiparticle continuum and the amplitude fluctuations of the SC order parameter, also named Higgs mode, are resonant at $2\Delta$. As a consequence, only a precise theoretical estimate  of the relative intensity of the two contributions, or the analysis of their dependence on the polarization of the pump field with respect to the main crystallographic axes, can be used to disentangle the origin of the $2\Delta$ resonance. 

Trying to understand which one, among the BCS and the Higgs contribution, is the main source for this nonlinear effects has been the subject of an intense experimental\cite{shimano_prl12,shimano_prl13,shimano_science14,shimano_prb17,giorgianni_natphys19,wang_natphot2019,wang_mgb2_prb21} and theoretical work\cite{shimano_science14,aoki_prb15,cea_prb16,aoki_prb16,cea_leggett_prb16,shimano_prb17,cea_prb18,giorgianni_natphys19,silaev_prb19,shimano_prb19,tsuji_prr20,seibold_prb21} in the last few years. In the attempt to resolve such a controversy, it turned out that despite both VIS and THz pulses can trigger collective excitations via an ISRS excitation process, one finds that in a superconductor the non-linear optical kernel $K$ controlling the response is {\em different} in the two cases, i.e.
\be
\lb{raman}
K^{THz}_{Raman}\neq K^{eV}_{Raman}.
\ee
In other words, even though in both cases only Raman-like excitations are involved, the microscopic fermionic processes mediating the light-mode coupling are different in the case of a difference-frequency (in the visible) or of a sum-frequency (in the THz) excitation. More specifically, while $K^{eV}_{Raman}$ in the widely used effective-mass approximation\cite{deveraux_review} essentially probes  the BCS continuum via lattice-modulated charge fluctuations, as given by diamagnetic-like coupling of electrons to light, $K^{THz}_{Raman}$ also probes the BCS continuum via paramagnetic-like coupling of the electron current to light. Such a difference arises once impurity effects are taken into account\cite{silaev_prb19,shimano_prb19,tsuji_prr20,seibold_prb21}, since current is no more conserved in the presence of disorder. Such a difference in turns leads to distinct selection rules for THz-driven or VIS-driven excitations, making eventually the Higgs-mode excitation, irrelevant in the clean limit\cite{deveraux_review,cea_prb16},  sizeable in the strong-disorder limit\cite{silaev_prb19,shimano_prb19,tsuji_prr20,seibold_prb21}. 

The above result is particularly relevant in the context of recent experiments in unconventional cuprate superconductors. So far, the only observation of marked $2\Delta$ oscillations in cuprates has been done using visible light\cite{carbone_pnas12}, and it has been ascribed to the BCS response, in agreement with the usual interpretation of conventional Raman measurements in cuprates\cite{deveraux_review,cea_prb16}. 
Nonetheless, the specular experiments performed with THz light have been almost exclusively interpreted so far as a response of the Higgs mode\cite{shimano_prl18,kaiser_natcomm20,shimano_prb20,chu_cm21}, despite the lack of a theoretical calculation of $K^{THZ}_{Raman}$ for these systems. The aim of the present manuscript is to provide such a quantitative analysis, by means of the direct computation of the non-linear THz Raman tensor within a realistic microscopic model for cuprates.   In particular, we will take advantage of the numerically-exact calculation in the presence of disorder recently discussed in Ref.\ \cite{seibold_prb21} to  provide a quantitative estimate of $K^{THZ}_{Raman}$ as a function of the band parameters and/or the doping level. As we shall see, for the band structure and disorder level of cuprates the contribution of the Higgs mode is still fairly subdominant with respect to the BCS response, whose polarization dependence can be well understood by taking into account both diamagnetic and paramagnetic processes at an appropriate disorder level.  While this result completely accounts for the polarization dependence observed for third-harmonic generation
experiments in transmission\cite{kaiser_natcomm20,chu_cm21}, where the signal is pretty much isotropic, it leaves nonetheless some open questions for the polarization observed via THz pump-optical probe experiments\cite{shimano_prl18,shimano_prb20}, that report instead a sizeable anisotropy of the signal for overdoped samples. We will argue below that such a difference can be ascribed to the peculiar role of paramagnetic-like processes when THz and VIS light pulses act simultaneously, leading in principle to an intermediate effect between the two extreme cases encoded in Eq.\ \pref{raman}. 

\section{Non-linear response from resonant modes}

\begin{figure}[h!]
\centering
  \includegraphics[height=6cm]{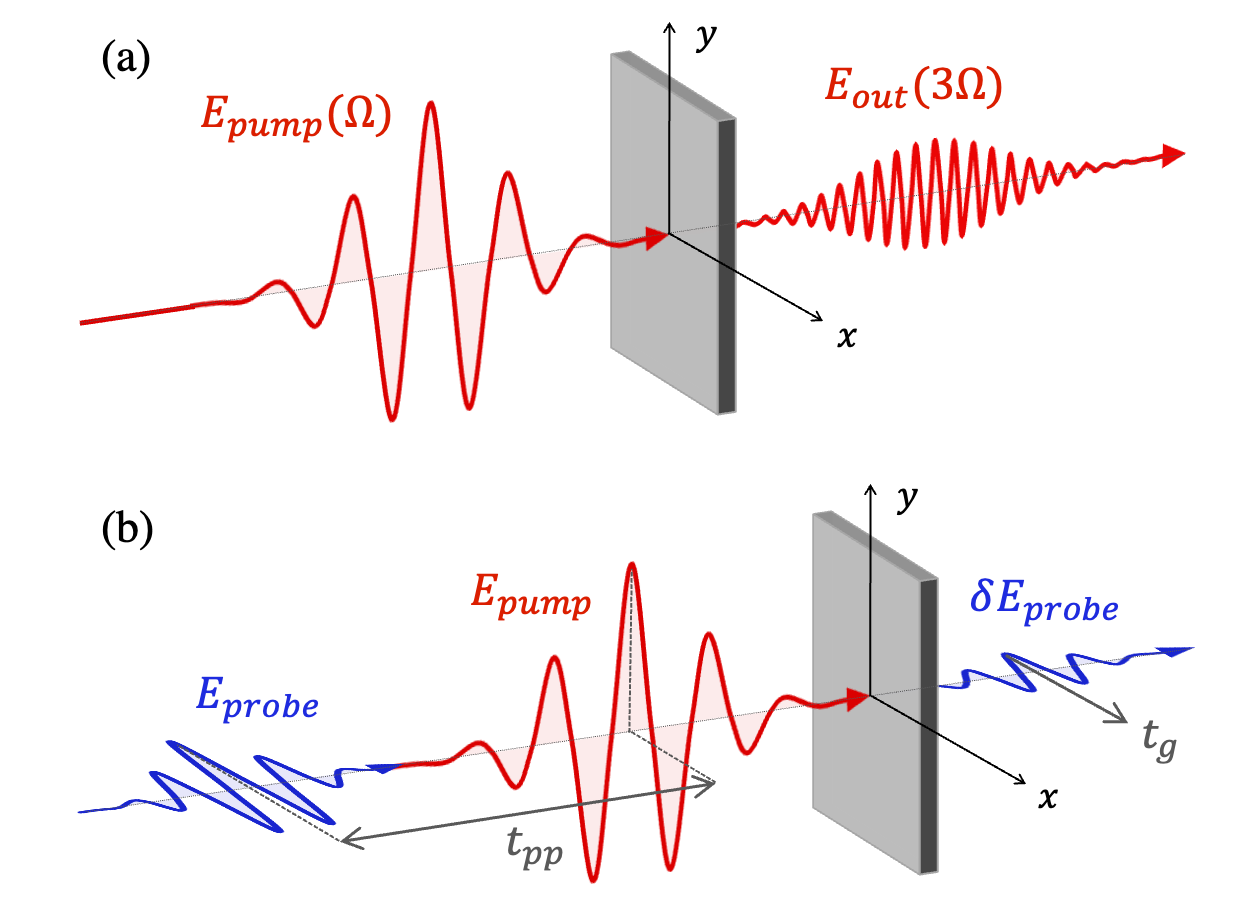}
  \caption{Schematics of unconventional THz spectroscopic techniques in transmission. In a typical THG experiment (a), the sample is perturbed using a multi-cycle (narrowband) THz pump pulse at $\O$ and one collects the induced transmitted component of the field at $3\O$. In pump-probe measurements (b), instead, one first perturbs the sample with an intense (multi-cycle or single-cycle) pump field and then records the differential transmitted component, with and without the pump, of a weak probe pulse, as a function of the time delay $t_{pp}$ between the two pulses at fixed observation time $t_{g}$. 
}
  \label{setup}
\end{figure}

Before discussing the specific case of cuprates, let us provide a general interpretative scheme to understand how the measured quantity in different unconventional THz spectroscopic techniques can be related to the same non-linear optical response, setting the basis for a more rigorous understanding of the experimental results. Since a detailed discussion has been already provided in Ref.\ \cite{udina_prb19}, here we will recast the main results only. In general, one can distinguish between two classes of experiments (see Fig.\ \ref{setup}): (a) measurements of third-harmonic generation (THG) in transmission and (b) pump-probe protocols, both in transmission and reflection configuration. In the former case one excites the sample with an intense THz laser pulse $E_{pump}(t)$ and records the transmitted electric field $E_{out}(t)$. These experiments are usually performed with multi-cycle laser pulses, such that the spectrum of the incoming radiation $E_{pump}(\omega)$ is strongly peaked around a central frequency $\omega =\Omega$. In this situation, the THG process manifests as significant spectral component of $E_{out}(\omega)$ at $3\Omega$, with different amplitudes according to the incoming $\Omega$ value or to the specific temperature of the sample. In the case (b), instead, the system can be excited either with a single-cycle or with a multi-cycle THz laser pulse. Single-cycle pulses last for less than 1 ps, and are thus associated with a relatively broad spectrum around the central frequency $\Omega$. The subsequent detection process occurs using a weak THz or NIR/VIS probe pulse. In both cases the probing field is delayed by $t_{pp}$ with respect to the pump laser pulse and the recorded signal is usually a differential change $\delta E_{probe}$ in the transmitted or reflected probe field measured with and without the pump, recorded as a function of $t_{pp}$. 
%
%

Following the field-theory approach developed e.g.\ in Refs.\ \cite{cea_prb16, giorgianni_natphys19,udina_prb19}, in order to reproduce the experimental findings one has to compute the third-order current flowing inside the sample, given in full generality by the partial derivative with respect to the external e.m.\ field of the fourth-order action $S^{(4)}$ written in terms of the e.m.\ vector potential $\bf A$, where ${\bf E} \equiv - \partial_t {\bf A}$, i.e.\ 
\be
\begin{gathered}
\lb{s4gen}
S^{(4)}[\bA]=e^4 \int d\O_1d\O_2 d\O_3 \sum_{\a \b \g \d}A_\a(\O_1)A_\b(\O_2) \times \\
\times K_{\a\b\g\d}(\O_1,\O_2,\O_3) A_\g(\O_3)A_\d(-\O_1-\O_2-\O_3),
\end{gathered}
\ee
with $e$ the electron charge and $K_{\a\b\g\d}$ a third-order tensor, which depends in the most general case on four spatial indexes and three incoming frequencies. While Eq.\ \pref{s4gen} accounts for all the possible third-order processes contributing to the nonlinear current, we will focus for the moment on the sub-set of processes allowing us to rewrite the effective action as
\bea
S^{(4)}[\bA]&=&\int d\O \sum_{\a \b \g \d} \bar A^2_{\a\b}(\O)K_{\a\b;\g\d}(\O)\bar A^2_{\g\d}(-\O)=\nn\\
\lb{s4res}
&=&\int dt dt' \sum_{\a \b \g \d} A^2_{\a\b}(t)K_{\a\b;\g\d}(t-t')A^2_{\g\d}(t'),\hspace{0.5cm}
\eea
where we defined $\bar A^2_{\a\b}(\O) \equiv \int d\o A_\a(\O)A_\b(\O-\o)$ as the Fourier transform of $A_\a(t)A_\b(t)$ and we put the semicolon between spacial indices in $K$ to underline that the nonlinear kernel has a Kubo-like structure, with two vertices carrying two field components each. The nonlinear current along the generic direction $\a$ then reads
\be
\lb{jnl}
J_\a^{NL}(t)=-2e^4 \sum_{\b \g \d} A_\b(t) \int dt' K_{\a\b;\g\d}(t-t')A_\g(t') A_\d(t').
\ee
When dealing with THG measurements only the $A_{pump}(t)$ field is present, meaning that, if light is shed along e.g.\ the $y$ crystallographic axis and one collects the transmitted field component at $3\Omega$ along the same direction (see Fig.\ \ref{setup}a), the resulting THG intensity is proportional to $\left | J_y^{NL}(3\Omega) \right|^2$, since the linear response to the incoming field does not contain additional harmonics. In particular, by approximating the incident multi-cycle incoming field with a monochromatic one $A_y(t) = A_0 \cos(\Omega t)$, the intensity of the THG signal can be written as 
\be
I^{THG}(\Omega) \sim \left | \int J_{y}^{NL}(t)e^{i3\Omega t} \right |^2 \propto |K_{yy;yy}(2\Omega) A_0^3 |^2,
\lb{Ithg}
\ee
showing that $I^{THG}$ scales as the squared modulus of the nonlinear kernel evaluated at {\em twice} the frequency of the incoming pump field.
\begin{figure}[h!]
\centering
  \includegraphics[width=8.2cm]{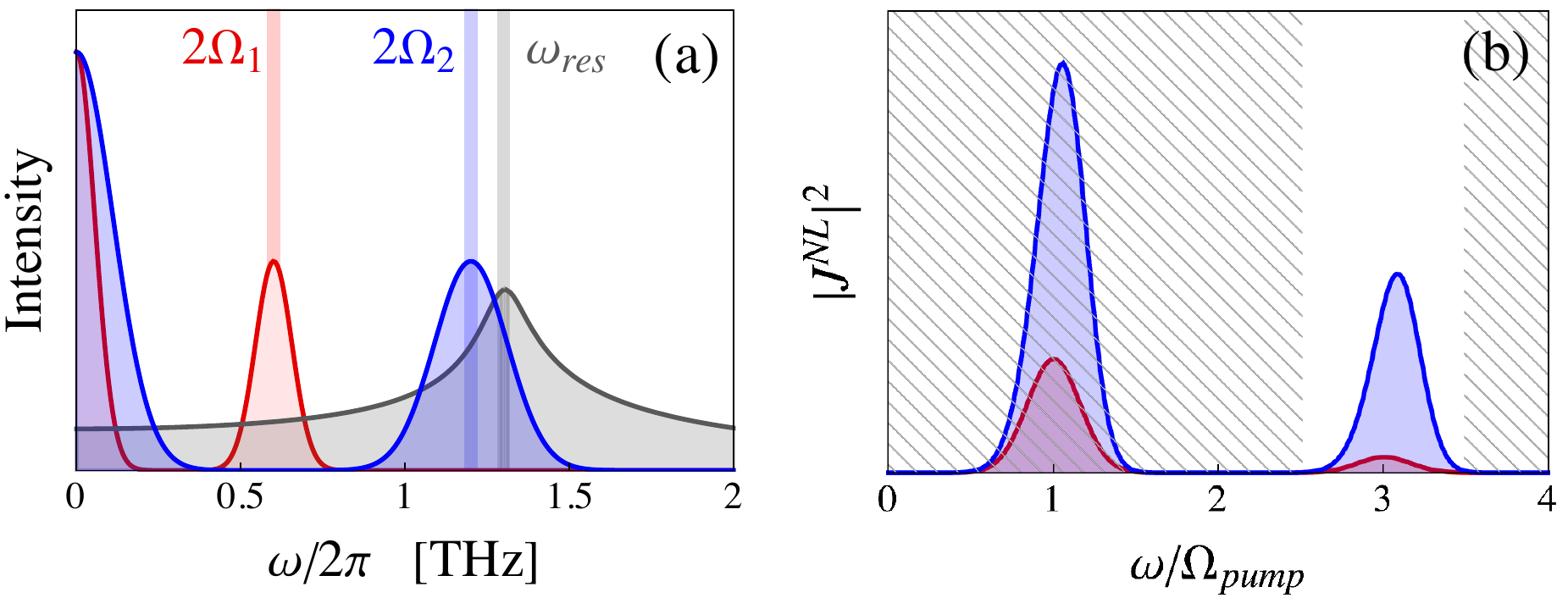}
  \caption{THG with a narrow-band pulse. (a) Spectral component of the non-linear kernel $K(\omega)$ resonant at $\omega_{res}/2\pi \simeq1.4$ THz (grey line) along with the squared pump pulse $\bar A^2(\omega)$ for two values of the central frequency, $\Omega_1/2\pi=0.3$ THz ($\tau=9.5$ ps, red line) and $\Omega_2/ 2\pi=0.6$ THz ($\tau=4.8$ ps, blue line). (b) Corresponding spectrum of the non-linear current $J^{NL}(\omega)$ computed using Eq.\ \pref{nonlin}. The dashed area represents the spectral region not relevant for the THG.}
  \label{narrow}
\end{figure}

For realistic multi-cycle band pulses the pump field can be described  by periodic oscillations convoluted with a gaussian decay \cite{udina_prb19}, i.e.\ $A_{y}(t)=A_0 e^{-(2t\sqrt{\ln2}/\tau_{p})^2} \cos(\Omega t)$. In this case the non-linear current $J^{NL}$ is given by the more general convolution\cite{cea_prb16,udina_prb19}
\be
J^{NL}_y(\omega) = -\int d\omega' A_y(\omega-\omega')K_{yy;yy}(\omega')\bar A_y^2(\omega').
\label{nonlin}
\ee
A typical spectrum of $J^{NL}(\omega)$ is shown in Fig.\ \ref{narrow}b, with the blue and red lines corresponding to two possible values of the central frequency $\Omega$ of the pump.  Here the non-linear kernel $K(\omega)$ is taken with a marked resonance at $\omega_{res}\simeq 1.4$ THz, and its form is the one expected for a superconductor (see below).  As one can see, since $\bar A^2(\omega)$ has spectral components both around $\omega \simeq 0$ and $\omega \simeq 2\Omega$, see Fig.\ \ref{narrow}a, the non-linear current $J^{NL}(\omega)$ has components around $\omega\simeq \Omega$, weighted approximately\cite{cea_prb16} with $2K(0)+K(2\Omega)$, and around $\omega\simeq 3\Omega$, weighted approximately with $K(2\Omega)$, see also Eq.\ \pref{Ithg} above.  As a consequence, the maximum value  of the intensity of $J^{NL}(\omega)$ around $\omega\approx 3\Omega$ is obtained when $\Omega\approx \omega_{res}/2$, so that the overlap of $\bar A^2(\omega\approx 2\Omega)$ with $K(\omega\approx \omega_{res})$ is the largest, see Fig.\ \ref{narrow}b. In other words, THG is maximized when two photons of the pump field resonantly excite the collective mode responsible for the divergence of $K(\omega)$. 

 \begin{figure}[h]
\centering
  \includegraphics[height=6.5cm]{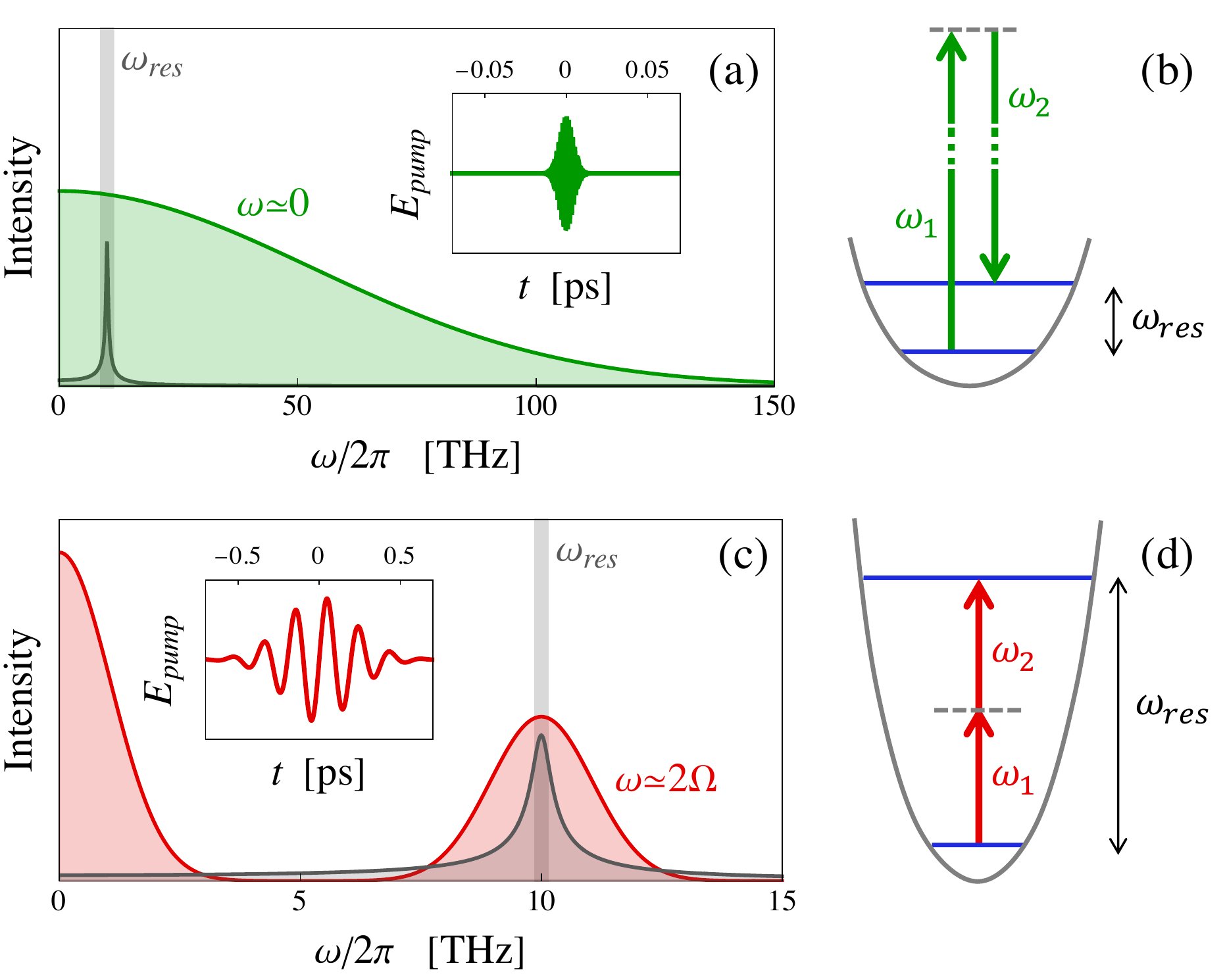}
  \caption{Difference-frequency vs sum-frequency excitations in pump-probe protocols, for a generic non-linear kernel $K(\omega)$ resonant at $\o_{res}$ (grey line). (a) When VIS pump pulses are applied (inset), the kernel overlaps with the $\omega \simeq 0$ peak in $\bar{A}_{pump}^2(\omega)$ (green line). (b) Two photons at energies $\omega_1, \omega_2$, taken from the (relatively broad) pump spectrum, lead to the difference-frequency excitation of the mode, i.e.\ $\omega_{res}=\omega_1-\omega_2$. (c) When THz pump pulses are applied (inset), the kernel overlaps with the $\o \simeq 2\O$ peak in $\bar{A}_{pump}^2(\omega)$ (red line). (d) The two $\omega_1, \o_2$ photons lead to the sum-frequency excitation of the mode at $\o_{res}=\o_1+\o_2$.}
  \label{dfe_sfe}
\end{figure}

The connection with THz sum-frequency two-photon processes is also evident in the typical experimental set-up behind pump-probe protocols. For what concerns this kind of experiments, indeed, the differential transmitted field $\delta E_{probe}$ can be related once more to the nonlinear current \pref{jnl}, where now both the pump and probe pulses, as well as their relative time delay $t_{pp}$, must be taken into account. In particular, by explicitly rescaling the pump field as $A_{pump}(t) = \tilde A_{pump} (t+t_{pp})$, in such a way that both $\tilde A_{pump}$ and $A_{probe}$ are centered around $t=0$, and by fixing the observation time at $t_{g}$, it can be easy shown\cite{giorgianni_natphys19,udina_prb19} that in the cross-polarized configuration depicted in Fig.\ \ref{setup}b the measured quantity reads
\be
\begin{gathered}
\lb{deltaE_tpp}
\delta E_{probe}(t_{pp}) \propto A_{probe}(t_g) \int dt'K_{xx;yy}(t_g+t_{pp}-t')\times \\
\times \left[\tilde A_{pump}(t')\right]^2.
\end{gathered}
\ee
Since the acquisition time $t_g$ is fixed, the previous expression shows that the probe field simply acts as a multiplying factor, setting the overall amplitude and phase of the oscillations, while the time-evolution of the signal is controlled by the convolution between the kernel and the squared pump field. More interestingly, if we Fourier transform Eq.\ \pref{deltaE_tpp} with respect to $t_{pp}$ we find a very compact expression for the power spectrum of the differential transmitted field:
\be
\lb{E_omega}
\delta E_{probe}(\o)\propto K_{xx;yy}(\omega)\bar{A}_{pump}^2(\omega).
\ee
Eq.\ \pref{E_omega} allows one to predict the presence of oscillations in $\delta E_{probe}(t_{pp})$ at the $\omega_{osc}$ frequency which dominates the convolution between the kernel and the squared pump pulse. In addition, Eq.\ \pref{E_omega} provides a simple way to understand what marks the difference between pump-probe experiments performed using a pump pulse with a central frequency $\Omega$ in the NIR/VIS or in the THz range. For the sake of clarity, let us focus again on a non-linear kernel $K(\omega)$ which displays a marked maximum at a given frequency $\omega_{res}$ in the THz range, see Fig.\ \ref{dfe_sfe}. 
 When dealing with NIR/VIS pulses $2\Omega \gg \omega_{res}$, so the only relevant overlap in Eq.\ \pref{E_omega} occurs between $K(\omega)$ and the $\omega \simeq 0$ peak of $\bar{A}_{pump}^2(\omega)$. This means that the mode is excited via a difference-frequency process, in which two photons taken from the pump have energies $\omega_1 \simeq \Omega$ and $\omega_2 \simeq \Omega - \omega_{res}$ (see Fig.\ \ref{dfe_sfe}a-b), in full analogy with stimulated Raman scattering. On the contrary, for THz pump pulses designed to have $2\Omega \sim \omega_{res}$, what matters in Eq.\ \pref{E_omega} is the overlap of the non-linear kernel with the $2\Omega$ peak in $\bar{A}_{pump}^2(\omega)$ and one is dealing with a sum-frequency excitation, where $\omega_{1,2} \simeq \omega_{res}/2$ (see Fig.\ \ref{dfe_sfe}c-d).  The basic mechanism is then the same already highlighted before, when discussing THG processes with multi-cycle THz pulses.
 
%
\begin{figure}[h]
\centering
  \includegraphics[height=9.cm]{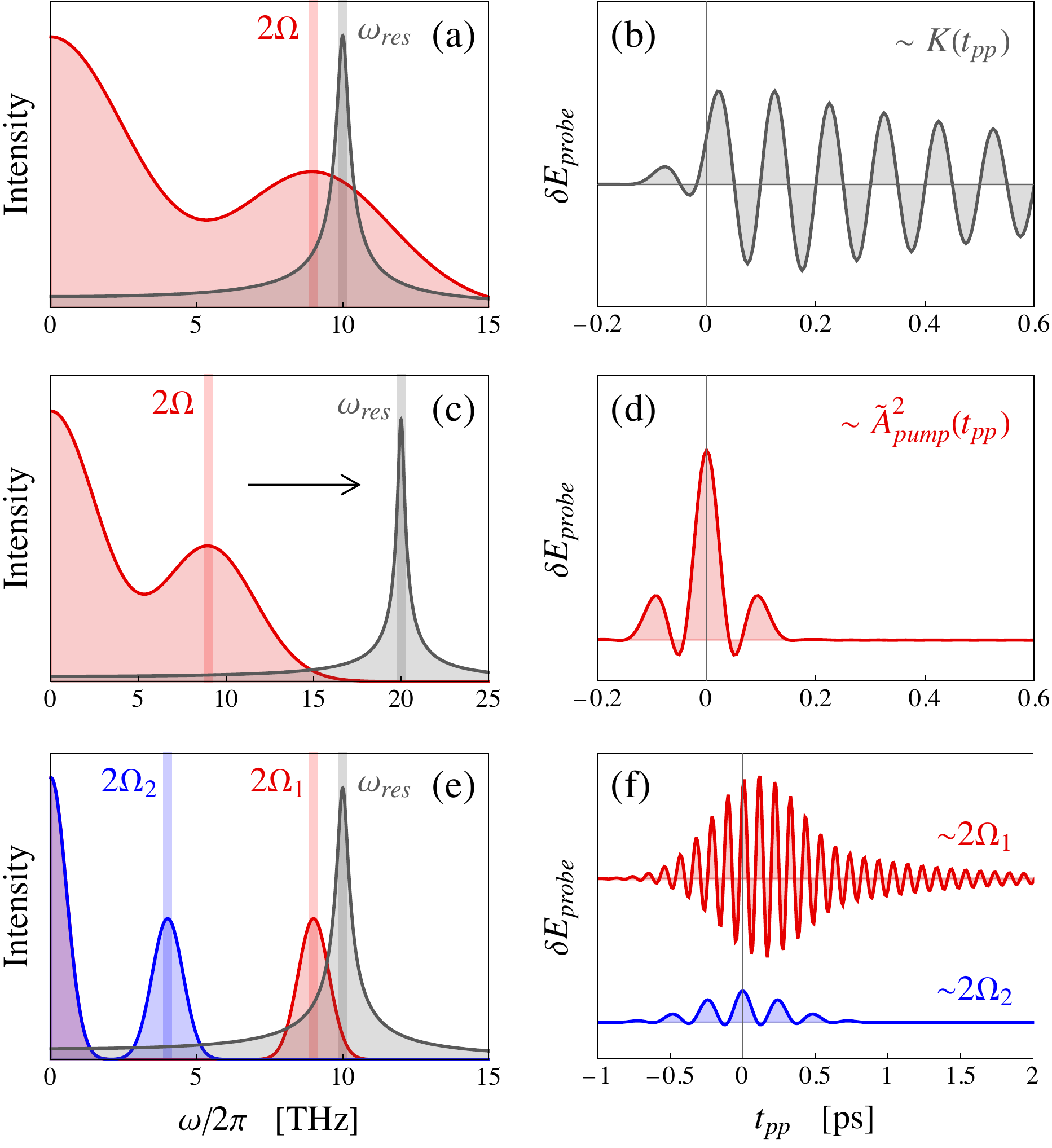}
  \caption{Pump-probe experiments with THz fields. When using single-cycle pump pulses, if the resonance in the nonlinear kernel (grey line) is close to $\omega = 2\Omega$ (a), the squared pump field ($\tau=0.2$ ps, red line) is well approximated by a constant in Eq.\ \pref{E_omega}, and the differential transmitted probe field $\delta E_{probe}(t_{pp})$ oscillates at $\omega_{res}$, following $K(t_{pp})$ in time (b). When instead $\omega_{res} \gg 2\Omega$ (c), the convolution is dominated by $\bar A_{pump}^2(\omega)$ and the differential probe field $\delta E_{probe}$ follows the squared pump field in the $t_{pp}$ time domain, leading to the so-called THz Kerr effect (d). Analogously, for multi-cycle pump fields only the response at $2\Omega$ dominates (e), and due to stronger duration of the pulse ($\tau_{1,2} = 1$ ps) one can clearly identify in $\tilde A^2(t_{pp})$ marked oscillations at $2\Omega$ (f), with an amplitude that is strongly enhanced when $\Omega\simeq \omega_{res}/2$ (red line) as compared to the case  when $\Omega \ll \omega_{res}/2$ (blue line). }
  \label{broad}
\end{figure}

Once established the general mechanism, it is useful to further distinguish between three different cases which determine the final time evolution of  the  $\delta E_{probe}(t_{pp})$ signal. Let us first consider the case of short single-cycle THz light pulses whose duration $\tau_p$ is much shorter  than the time scale $\tau_{res}$ of the $\omega_{res}$ resonance, as set by the inverse width of the $\omega_{res}$ peak in the non-linear kernel $K(\omega)$, i.e. $\tau_p<\tau_{res}$, and 
 having a central frequency  $2\Omega \simeq \omega_{res}$. In this case, that is usually named as "quench",  one finds that $\bar A_{pump}^2(\omega)$ is rather flat around $\omega_{res}$ (Fig.\ \ref{broad}a) and the spectrum of the differential transmitted field \pref{E_omega} essentially follows the optical kernel, i.e.\ $\delta E_{probe}(\omega) \sim K(\omega)$. Notice that this condition is also fulfilled for eV light pulses, since their typical duration $\tau_{p} \sim10$ fs  makes $\tau_p\ll \tau_{res}$. As a consequence, one still obtains that $\bar A^2(\omega\approx 0)$ is very flat  around the resonance, see Fig.\ \ref{dfe_sfe}a, and consequently $\delta E_{probe}(\omega) \sim K(\omega)$. In both cases the main outcome is that in the time domain the differential electric field is expected to oscillate at the resonance frequency of the kernel, i.e. $\omega_{osc}\approx \omega_{res}$, see Fig.\ \ref{broad}b.
 Notice also that for THz light pulses the possibility to observe $\omega_{res}$ oscillations strongly relies not only on the antiadiabatic condition $\tau_p<\tau_{res}$  but also on the matching condition $2\Omega \simeq \omega_{res}$ of the pulse central frequency. Indeed, when the pump field moves out of resonance, as it happens e.g. for $\Omega\ll \omega_{res}$, the only relevant overlap in Eq.\ \pref{E_omega} is between $\bar A^2(\omega\approx 0)$ and $K(\omega\simeq 0)$,  leading to a nearly instantaneous contribution in $\delta E_{probe}(t_{pp})$ following the squared pump field in time, also referred to in the literature as THz Kerr effect (Fig.\ \ref{broad}d). An analogous effect is found when the system is driven by a narrow multi-cycle pump pulse, such that $\tau_p>\tau_{res}$. 
Such a condition, that is usually named as a "drive" pulse, translates in a $\bar A_{pump}^2(\omega)$ narrower than the $\omega_{res}$ peak in $K(\omega)$, see Fig.\ \ref{broad}e. As a consequence, the convolution \pref{E_omega} is dominated by the spectral components of $\bar A_{pump}^2(\omega)$ and $\delta E_{probe}(t_{pp})$ is once more proportional to the squared field in time, oscillating at $\omega_{osc}\approx 2\Omega$, i.e. at {\em twice} the central frequency of the narrow-band light pulse, see Fig.\ \ref{broad}f. 


\section{Superconducting modes and the role of disorder}

%
The results discussed in the previous Section are rather general, since they only require the existence of a marked resonance in the non-linear kernel $K(\omega)$ at a given frequency $\omega_{res}$, whatever is its origin. Thus, the same scheme can be used to understand non-linear excitation of phonons as well as of electronic collective modes\cite{udina_prb19}, as they emerge across a phase transition to a SC or a charge-density-wave (CDW) state. Let us then see how one can interpret the experiments in conventional superconductors, like e.g.  NbN\cite{shimano_prl12,shimano_prl13,shimano_science14,shimano_prb17}, Nb$_3$Sn\cite{wang_natphot2019} and MgB$_2$\cite{giorgianni_natphys19,wang_mgb2_prb21}, on the light of such a general paradigm. THG experiments have been performed in disordered NbN\cite{shimano_science14} by fixing the central frequency $\Omega$ of a multi-cycle pulse while changing the temperature of the sample. The general result is an enhancement of the THG below $T_c$, with a maximum at the temperature where $\Omega=\Delta(T)$. In the light of the previous discussion, see Eq.\ \pref{Ithg} and Fig.\ \ref{narrow}, this implies that the SC non-linear optical kernel $K_{SC}(\omega)$ has a maximum at $\omega_{res}=2\Delta(T)$. Analogously, pump-probe protocols with the same multi-cycle pulse have reported\cite{shimano_science14}  $2\Omega$ oscillations, as for the case shown in Fig.\ \ref{broad}f. When instead the system is quenched with a single-cycle THz pulse\cite{shimano_prl12,shimano_prl13}, the $\delta E_{probe}(t_{pp})$ signal shows oscillations at $\omega_{osc}=2\Delta$, leading again to a kernel $K_{SC}(\omega)$ resonant at $2\Delta$, as explained while discussing Fig.\ \ref{broad}b above. When other collective modes are present, as it is the case for e.g.\ the Leggett mode in MgB$_2$, one can selectively see the resonance at twice the gap\cite{wang_mgb2_prb21} or at the Leggett-mode frequency\cite{giorgianni_natphys19} by tuning the pump frequency. In the latter case, it has been also tested\cite{giorgianni_natphys19}  the strong increase of $2\Omega$ oscillations in the pump-probe signal when $\Omega\simeq \omega_{res}/2$, as shown in Fig.\ \ref{broad}f. 

\begin{figure}[h]
\centering
  \includegraphics[width=8.2cm]{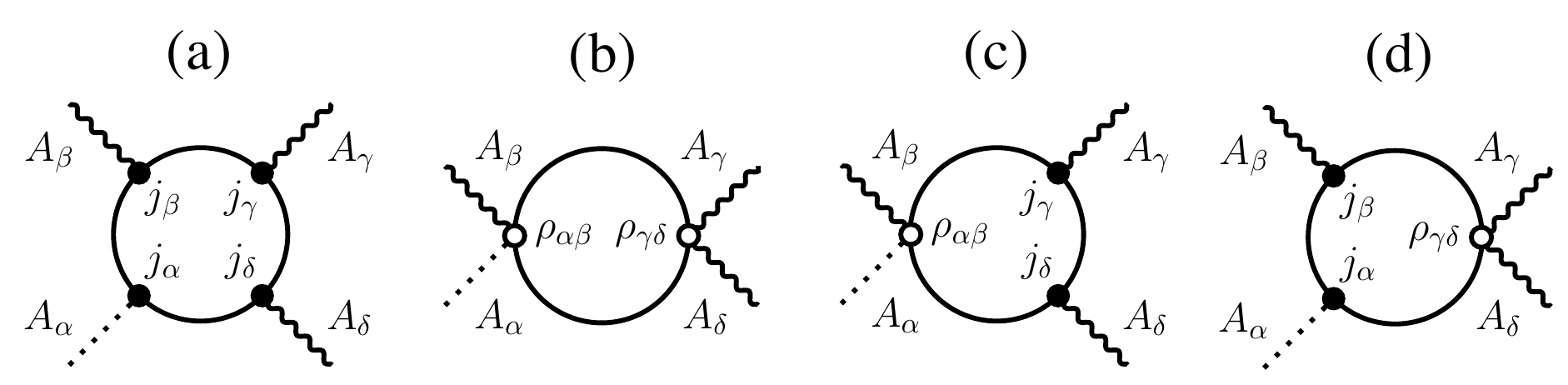}
  \caption{Schematic of some of the Feynman diagrams contributing to the non-linear optical response in an electronic systems. Here solid lines denotes electronic Green's functions and wavy lines the e.m. gauge field, $\alpha,\beta, \gamma, \delta$ being spatial indexes. The dotted line denotes the derivative with respect to one field component, as needed to compute the current according to Eq.\ \pref{jnl}. (a) Paramagnetic-like processes are built with current-like e.m.\ field insertions, denoted by a full circle. (b) Diamagnetic-like processes are built with density-like e.m.\ field insertions, denoted by an empty circle. Diagrams with two diamagnetic vertexes have the form of Kubo-like response functions. (c)-(d) Mixed terms.  }
  \label{diagrams}
\end{figure}

Focusing now on the case of NbN, the experimental findings are all consistent with a SC non-linear optical kernel $K_{SC}(\omega)$ resonant at $\omega_{res}=2\Delta(T)$. However, as mentioned in the Introduction, there has been for a while a debate in the literature about the identification of the relevant excitations responsible for such a resonance. Since this issue has been discussed at length in several manuscripts\cite{cea_prb16,aoki_prb16,cea_leggett_prb16,shimano_prb17,cea_prb18,silaev_prb19,shimano_prb19,tsuji_prr20,seibold_prb21,shimano_review19}, here we just summarize the main points in a schematic way. On very general grounds, to derive the non-linear kernel appearing in Eq.\ \pref{s4gen} one needs to integrate out all the electronic degrees of freedom of a model system where interacting electrons are coupled to the external gauge field $\bA$. Let us start from the case of a clean system. Assuming that interaction terms are gauge-invariant, the gauge field only couples to the kinetic term of the Hamiltonian, so one can expand the Hamiltonian $H(\bA)$ in the presence of the e.m. field as:
\be
\label{hA}
H(\bA)\simeq H(\bA=0)+{\bf j}\cdot \bA+ \rho_{\alpha\beta} A_\alpha A_\beta+{\cal O}(A^3),
\ee
where ${\bf j}$ is the electronic paramagnetic current and $\rho_{\alpha\beta}$ simply reduces to $\delta_{\alpha\beta}n/m$ for free electrons, where $n$ is the electron density and $m$ the mass. Notice that in a lattice model the simple expression \pref{hA} should be extended to include all the bands describing electrons moving in a periodic potential. To derive $S^{(4)}[\bA]$ one needs to sum over all possible electronic processes,  leading to a response of order $A^4$. In terms of the usual Feynman-diagrams expansion, this leads to all possible electronic loops with four external e.m.\ legs, as exemplified in Fig.\ \ref{diagrams}. In general, terms of order $A^4$ can be obtained e.g.\ with correlation functions including four paramagnetic-like terms ${\bf j}\cdot \bA$  or two diamagnetic-like terms $\rho_{\alpha\beta} A_\alpha A_\beta$ from Eq.\ \pref{hA}, plus all possible mixed combinations. When dealing with high-energy NIR/VIS light pulses interband transitions should be also considered, as for the usual Raman response. In the case of non-resonant excitations, a very popular description for the overall non-linear response of metals relies on the so-called effective-mass approximation\cite{deveraux_review}, that reduces the sum of {\em all} possible interband processes to the computation of lattice-modulated density fluctuations for the band at the Fermi level:
\be
\label{keV}
K^{eV}_{Raman}=\langle \rho^{i,s}_R \rho^{i,s}_R \rangle,
\ee
where $\rho^{i,s}_R=\sum_{\bk,\sigma} \gamma^{i,s}(\bk)c^{\dagger}_{\bk,\s}c_{\bk,
\sigma}$ is a lattice-modulated density-like electronic operator, $c^{(\dagger)}_{\bk,\sigma}$ being the annihilation (creation) operator for an electron with momentum $\bk$ and spin $\sigma$. The form factor $\gamma^{i,s}(\bk)$ is proportional to the momentum derivatives of the band dispersion $\epsilon(\bk)$ along the crystallographic  axes, in a combination dictated by the polarization of the incident ($i$) and scattered ($s$) light. Within this scheme, that has been widely used in the past to interpret Raman experiments in cuprates\cite{deveraux_review}, the enhancement of the non-linear response at $2\Delta$ in the SC state is simply a consequence of the fact that the density-like response evaluated at BCS level  probes the quasiparticle continuum at $\bq=0$, that is pushed  at $2\Delta$ below $T_c$. On the other hand, the Higgs response appears as a vertex corrections of such a density-like response in the amplitude channel, and it turns out to be quantitatively negligible\cite{cea_prb16,cea_prb18}, since amplitude fluctuations are weakly coupled to density fluctuations in the particle-hole symmetric BCS case\cite{cea_prb16,deveraux_review}. In short, whenever the non-linear kernel $K_{SC}(\omega)$ reduces to a density-like response as in the case of Eq.\ \pref{keV}, the Higgs contribution is quantitatively subdominant with respect to the BCS one. 

In the case of experiments with a THz driving pump field a similar approach has been followed so far. Since in this case the energy of the field is comparable to intraband transitions, the non-linear response has been derived by coupling the gauge field directly to the electronic band at the Fermi level, via a Peierls-like substitution in the kinetic term of the Hamiltonian\cite{shimano_science14,aoki_prb15,cea_prb16,aoki_prb16,shimano_prb17,cea_prb18,silaev_prb19,tsuji_prr20,seibold_prb21}. The expansion of the Hamiltonian then leads again to a structure similar to Eq.\ \pref{hA}, provided that the diamagnetic term is replaced directly with a density-modulated Raman operator $\rho_{R,\alpha\beta}=\sum_{\bk,\sigma} \Gamma_{\a\b} (\bk)c^{\dagger}_{\bk,\s}c_{\bk,
\sigma}$, where the modulation prefactor $\Gamma_{\a\b}(\bk)\sim \pd^2 \e_\bk/\pd k_\alpha k_\beta$ depends on the direction of the applied field with respect to the crystallographic axes. In the clean case only diamagnetic-like processes are relevant\cite{cea_prb16,cea_prb18},  so the non-linear kernel $K^{THz,clean}_{Raman}$ has the same behavior of $K^{ev}_{Raman}$ and it is dominated by the BCS response:
\be
\label{kthzclean}
K^{THz,clean}_{Raman}=\langle \rho_R \rho_R \rangle.
\ee
However, as shown by several authors in the very last years\cite{silaev_prb19,shimano_prb19,tsuji_prr20,seibold_prb21}, once {\em disorder} is taken into account the paramagnetic current is no more conserved at BCS level and also contributions mediated by the paramagnetic-like coupling term ${\bf j}\cdot \bA$ in Eq.\ \pref{hA} are finite,  becoming quantitatively larger than diamagnetic ones already at intermediate disorder level.  As a consequence, for a realistic SC system the THz response is controlled by intraband excitations mediated by both diamagnetic and paramagnetic correlation functions, i.e.
\be
\label{kthzdirty}
K^{THz,dirty}_{Raman}\sim \langle \rho_R \rho_R \rangle +\langle j_P j_P j_P j_P \rangle,
\ee
where the precise dependence on frequencies and spatial indexes has been omitted for simplicity. As far as the Higgs contribution is concerned, it turns out that paramagnetic-like processes can mediate a sizeable coupling of the e.m.\ field to Higgs fluctuations, such that at strong disorder the Higgs contribution can even dominate over the BCS one.  For thin films of conventional NbN it seems now plausible to ascribe the $2\Delta$ resonance of the non-linear kernel to an excitation of the Higgs mode\cite{silaev_prb19,shimano_prb19,tsuji_prr20,seibold_prb21}. However, cuprates are much cleaner system, and preliminary studies in Ref.\ \cite{seibold_prb21} suggest that the BCS response is still dominant for this level of disorder, and that collective SC phase-density fluctuations can give a bigger contribution to the non-linear response, as compared to the Higgs one. In the next Section we will see how detailed quantitative calculations with a realistic band structure support a predominant role of the BCS response. 


\section{Non-linear response in cuprates}

\begin{figure}[h]
\centering
\includegraphics[width=8.2cm,clip=true]{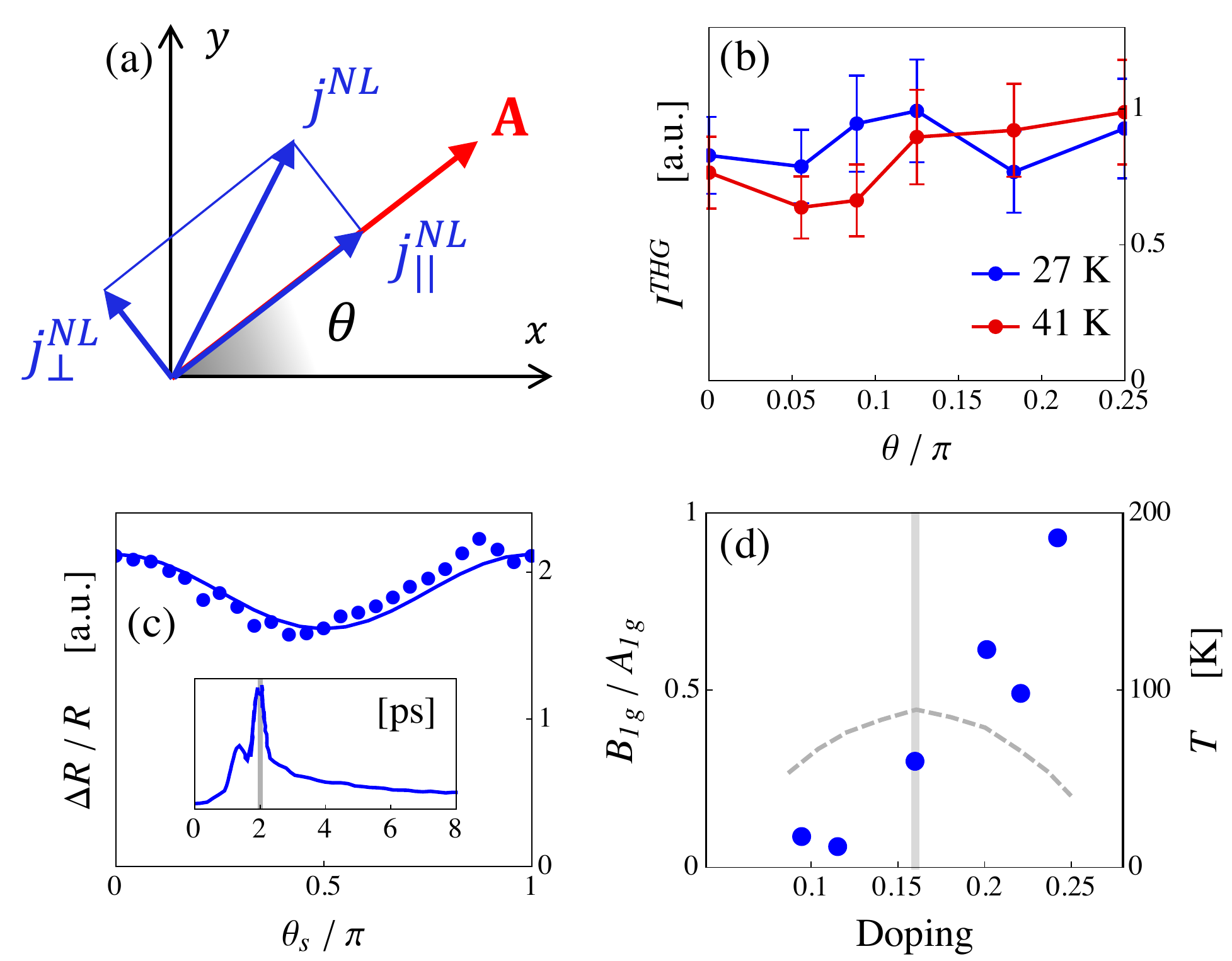}
  \caption{(a) Definition of the polarization dependence for THG experiments in cuprates, with the square lattice formed by Cu atoms in the CuO$_2$ planes. For a generic angle $\theta$ between the applied field $\bA$ and the $x$ direction, the measured THG response is in the field direction, so it depends on the component $j^{NL}_\pll$ of the non-linear current in the  $\theta$ direction. (b) THG results for a optimally doped La1$_{1.84}$Sr$_{0.16}$CuO$_4$ ($T_c$ = 45 K) sample at two temperatures below $T_c$, taken from Ref.\ \cite{kaiser_natcomm20}. (c) Polarization dependence of THz-Kerr effect in an optimally-doped Bi2212 compound ($T_c$=90 K) from Ref.\ \cite{shimano_prl18}. The inset shows the measured instantaneous response at a function of $t_{pp}$, while the main panel shows the recorded intensity at $t_{pp}$= 2 ps (marked by a vertical grey bar in the inset) as a function of $\theta_S$ for $\theta_P=0^{\circ}$. According to Eq.\ \pref{eq:js} a $K_{B1g}$ component appears as a minimum at $\theta_s=\pi/2$. (e) Doping dependence of the ratio $K_{B1g}/K_{A1g}$ in Bi2212 as a function of doping, taken from Ref.\ \cite{shimano_prl18}. The dashed gray line denotes the corresponding $T_c$ values, referring to the right vertical axis. The gray vertical line indicates the optimal doping.}
  \label{polarization}
\end{figure}
So far, THz-induced non-linear response in cuprates has been investigated via THG experiments in transmission\cite{kaiser_natcomm20, chu_cm21}, focusing on several classes of materials, and by means of pump-probe protocols aimed at measuring the THz Kerr effect in Bi-based compounds as a function of doping\cite{shimano_prl18,shimano_prb20}. In analogy with previous work in conventional superconductors, the experiments are performed by varying the temperature at fixed pump frequency. While all experiments clearly show a strong enhancement of the non-linear response below $T_c$, that is then naturally ascribed to the SC phase transition, the present data do not allow one to clearly identify a resonance of $K_{SC}(\omega)$ at twice the gap maximum. This is in part understood by noting that in cuprates the pump pulse has a central frequency much smaller that twice the gap. More specifically, a multi-cycle pulse with central frequency $\Omega= 0.7$ THz has been used for THG measurements\cite{kaiser_natcomm20,chu_cm21}, and a single-cycle THz pump field with central frequency $\Omega=0.6$ THz, and a probe in the VIS, have been used for the THz Kerr effect\cite{shimano_prl18,shimano_prb20}. In both cases the pump frequency ($\simeq 3$ meV) is much smaller than the gap value at $T=0$, that lies around 10-20 meV in cuprates. As a consequence, the resonance condition $\Omega=\Delta(T)$ only occurs very near to $T_c$, making it difficult in general to observe the resonant enhancement of the non-linear response. In addition, $d$-wave symmetry of the order parameter can in part smear out the divergence of the response at $2\Delta$, with $\Delta$ gap maximum, in analogy with what observed in Raman\cite{deveraux_review}. On the other hand, polarization results are rather robust, and the analysis of their behavior deep in the SC phase can help disentangling the nature of the relevant modes involved in the non-linear response.


For what concerns THG experiments, one usually measures the third-harmonic emission in the field direction by changing the angle $\theta$ that the pump field forms with the in-plane $x$ crystallographic axis, see Fig.\ \ref{polarization}, where the $xy$ plane represent the CuO$_2$ plane of cuprates. Since $I^{THG}\propto |J^{NL}(3\Omega)|^2$, see Eq.\ \pref{Ithg}, the relevant quantity to be computed is the non-linear current $j^{NL}_{\pll}$ in the field direction. The experimental results of Ref.\ \cite{kaiser_natcomm20, chu_cm21} are pretty much isotropic, at least within the error bars, see e.g. the data in Fig.\ \ref{polarization}b, taken from Ref.\ \cite{kaiser_natcomm20}. For THz pump-optical probe experiments one can vary both the pump $(P)$ and probe $(S)$ angles $\theta_{P,S}$ with respect to the crystallographic axes, and study the angular-dependence of the time-resolved modulation at a fixed $t_{pp}$. As we explained above, at low $T$ one is always in the condition $\Omega\ll \omega_{res}$, so the time-dependent response is expected to scale in the $t_{pp}$ domain with the square of the pump field, see Fig.\ \ref{broad}d, as indeed observed experimentally, see inset in Fig.\ \ref{polarization}c. However, in contrast to the case of THG experiments, THz Kerr effect measurements reveal a modulation of the signal (see  Fig.\ \ref{polarization}c) that increases as the doping increases, see Fig.\ \ref{polarization}d. In Ref.\ \cite{shimano_prl18} such a modulation has been analyzed in terms of a decomposition of $j^{NL}_\pll$ in the symmetry projections for the $D_{4h}$ point group relevant for cuprates:
\begin{equation}
\begin{gathered}
\label{eq:js}
j^{NL}_{\parallel}(\theta_S,\theta_P)=K_{A1g}+K_{B1g}\cos{2\theta_S}\cos{2\theta_P}+\\
+K_{B2g}\sin{2\theta_S}\sin{2\theta_P}.
\end{gathered}
\end{equation}
As it has been discussed in Ref.\ \cite{seibold_prb21}, such a decomposition is meaningful when  the non-linear kernel $K_{\alpha\beta\gamma\delta}$ only admits diamagnetic Kubo-like diagrams, that is not necessarily the case in the presence of disorder. Nonetheless, by preserving the notation of Ref.\ \cite{shimano_prl18} based on the decomposition \pref{eq:js}, the analysis of the experimental data gives $K_{B2g}=0$ and a ratio $K_{B1g}/K_{A1g}$ increasing with doping from almost 0 in underdoped samples to almost 1 in overdoped ones, see Fig.\ \ref{polarization}d. It is worth noting that using the same decomposition also for THG measurements in transmission one would obtain that
\begin{equation}
\label{eq:fit}
j_{\parallel}^{NL}(\theta)=K_{A1g}+K_{B1g}\cos^2(2\theta)+K_{B2g}\sin^2(2\theta).
\end{equation}
As a consequence, the isotropy of the THG signal in cuprates, see Fig.\ \ref{polarization}b, points to the existence only of the $K_{A1g}$ component, as seen at low doping in pump-probe experiments, see Fig.\ \ref{polarization}d, but making it puzzling the lack of a strong $B_{1g}$ component around optimal and overdoping, as  we will comment further in what follows. 

All the experimental findings summarized so far help disentangling the contribution to the non-linear response coming from the BCS response or from the other collective modes. Indeed, these results can be understood from the general scheme outlined in Sec.\ 2 by assuming that a sizeable non-linear kernel $K_{SC}(\omega)$ emerges below $T_c$, with a peculiar polarization dependence. 
To clarify the nature of the non-linear response in cuprates, we then study an extended version of the attractive Hubbard model that we used in Ref.\ \cite{seibold_prb21}, accounting for the effects of disorder and doping on a realistic band structure. More specifically we start from the electronic Hamiltonian
\begin{equation}\label{eq:hub}                                         
H=-\sum_{ij \sigma}t_{ij}c^\dagger_{i\sigma}c_{j\sigma} - |U|\sum_{i}n_{i\uparrow}n_{i\downarrow} +\sum_{i\sigma}V_i n_{i\sigma},                                      
\end{equation}
where the local potential $V_i$ is taken from a flat distribution $-V\le V_i \le +V$. We set $t_{ij}=t$ for the nearest neighbors hopping and a different $t_{ij}=-t^{\prime}$ for the next-nearest neighbors hopping to better reproduce the Fermi surface of cuprates. For numerical reasons, calculations are performed on a $32\times 32$ lattice, and we set $U/t=1.4$. This is the lowest value for which finite size effects can be neglected. For a nearest-neighbor hopping $t\sim200-300$ meV and a next-nearest-neighbors hopping of $t^{\prime}/t\sim-0.2$, one obtains gap values $\Delta \sim25-55$ meV compatible with ARPES measurements in overdoped Bi-2212 close to the nodal region, where pseudogap effects should be small \cite{shimano_prl18}. We consider the doping levels $p=0.1$ for the underdoped regime and $p=0.2$ for the overdoped. Notice that the polarization dependence, which is the main focus here, depends specifically on the band structure and disorder level, so even though the model \pref{eq:hub} does not include the $d$-wave symmetry of the order parameter it can nonetheless provide a good quantitative account of the realistic situation, especially at pump frequencies far from the resonance.

\begin{figure}[h]
\centering
  \includegraphics[height=6.5cm]{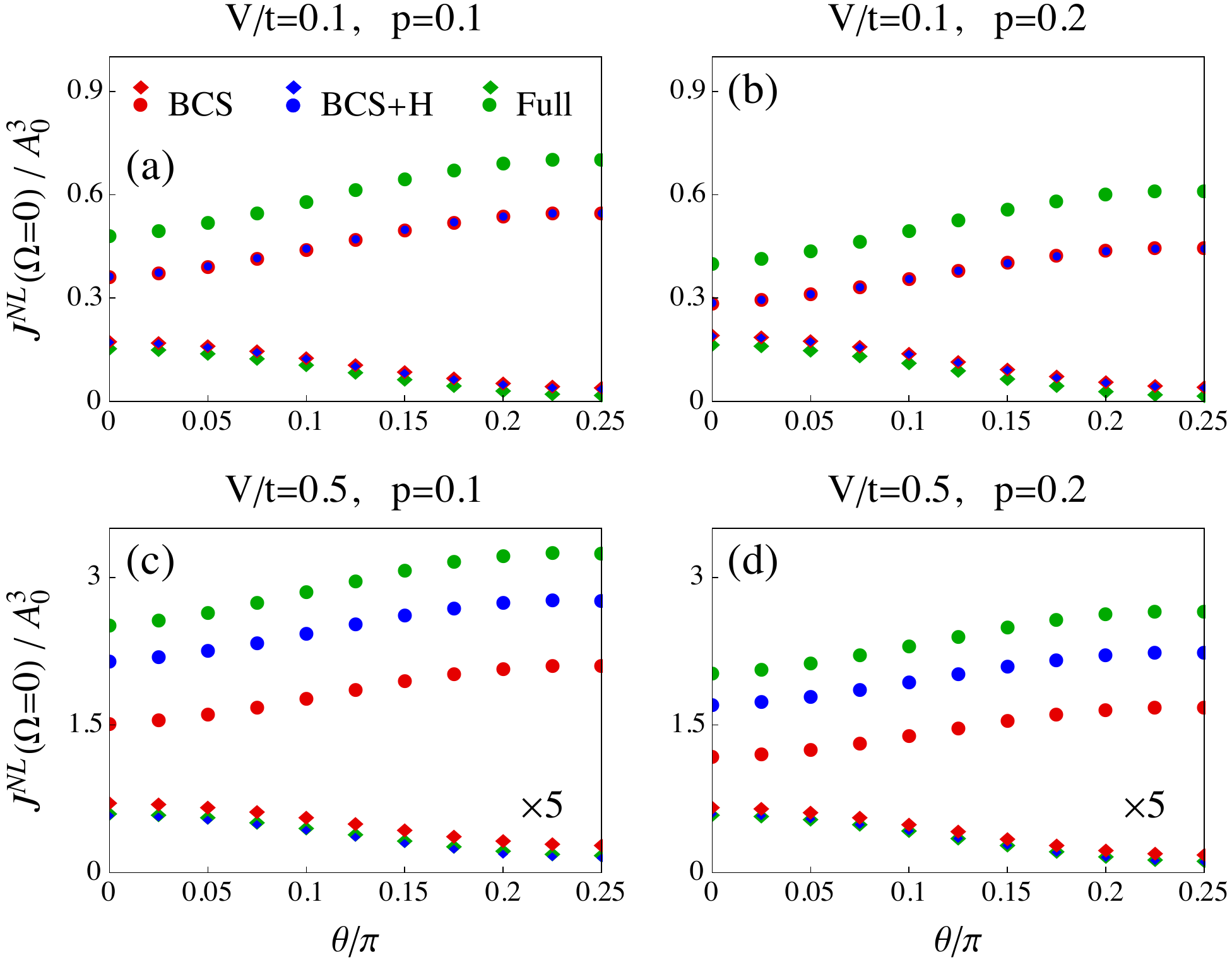}
  \caption{Numerical results for the non-linear current $j^{NL}_\pll(\theta)$ at $\Omega=0$ as a function of the angle $\theta$ for $V/t=0.1$ (upper row) and $V/t=0.5$ (lower row) at two doping levels. Diamonds refer to the diamagnetic contribution and circles to the paramagnetic one. "BCS" labels the pure BCS response, probing the quasiparticle continuum, "BCS+Higgs" the response including vertex correction in the SC amplitude (Higgs) channel, and "Full" the results obtained by including vertex corrections in all channels (SC amplitude, SC phase and charge). For more details on the notation see Ref.\ \cite{seibold_prb21}.  }
  \label{all_data}
\end{figure}

In order to establish a close connection with the experiments, we compute the transport scattering rate $\tau$ for each doping and disorder level, following the procedure outlined in Ref.\ \cite{seibold_prb21}. Experimental data in Ref.\ \cite{shimano_prl18} indicate a disorder level $\gamma/2\Delta \sim 0.85$, that is intermediate between our $V/t=0.1$, corresponding to $\gamma/2\Delta \sim0.03$ $(0.04)$ for $p=0.1$ $(p=0.2)$, and $V/t=0.5$ corresponding to $\gamma/2\Delta \sim1.12$ $(1.79)$ for $p=0.1$ $(p=0.2)$. By following the time evolution of the mean-field density matrix stemming from Eq.\ (\ref{eq:hub}), we calculate the third-harmonic current by selectively including the charge, phase and amplitude fluctuations, distinguishing between the paramagnetic and diamagnetic processes. We then focus on the zero-frequency value of the non-linear current, as a good approximation in the out-of-resonance condition. Further details on the numerical procedure have been reported in Ref.\ \cite{seibold_prb21}. 

\begin{figure}[h]
\centering
  \includegraphics[height=6.5cm]{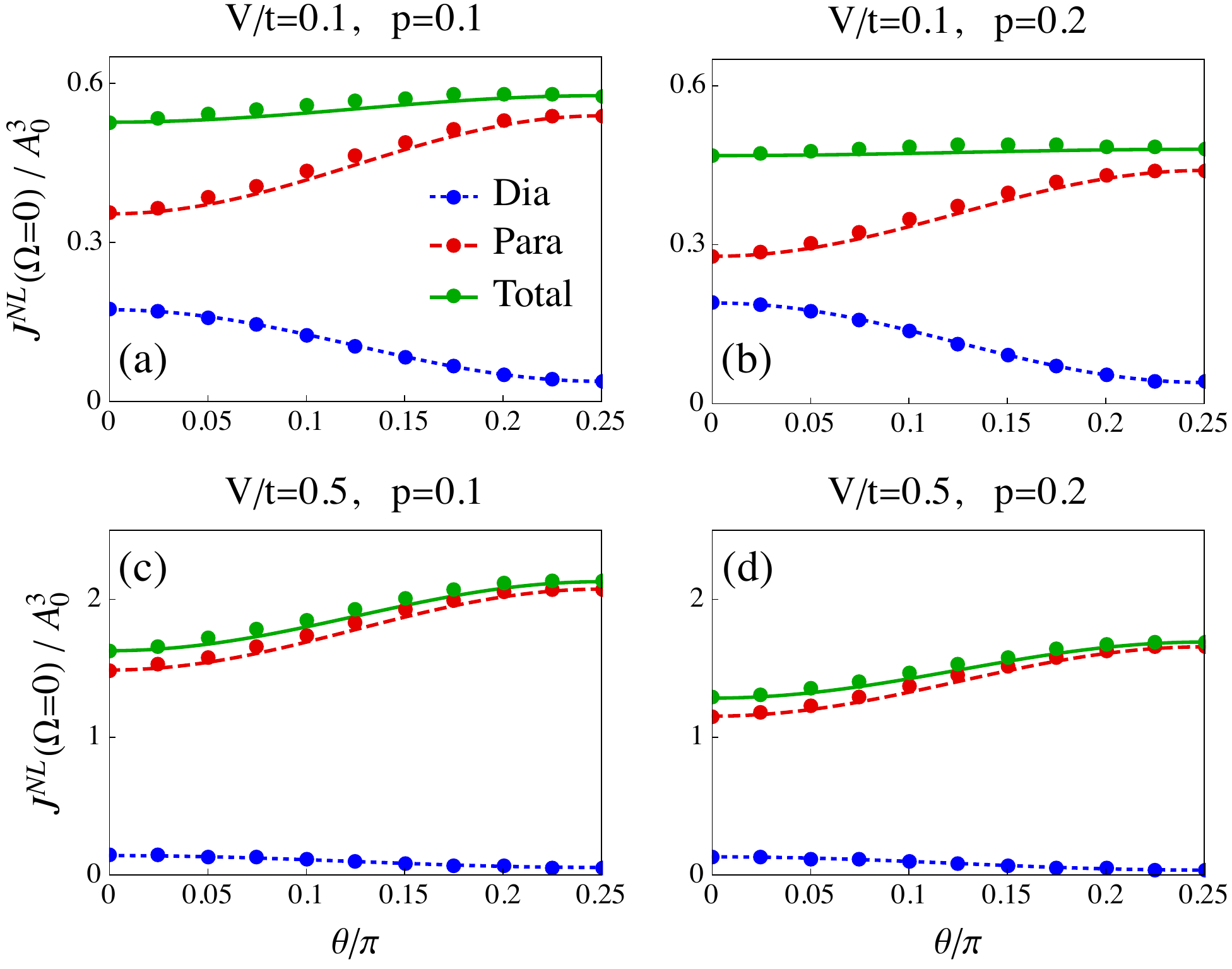}
  \caption{Polarization dependence of the BCS contribution to the non-linear current at $\Omega=0$ as given by diamagnetic (left) or paramagnetic (right) diagrams, for two different levels of doping and disorder. Solid lines represent a fit of the diamagnetic term with $K_{A1g}+K_{B1g}$ contributions, and of the paramagnetic one with $K_{A1g}+K_{B2g}$ contributions. The parameters of the fit are reported in Tables \ref{tab:table2}-\ref{tab:table3}. }
  \label{polariz}
\end{figure}

Fig.\ \ref{all_data} shows the $j_{\parallel}^{NL}(\theta)$ component of the nonlinear current as a function of the angle $\theta$, where we keep separate the BCS contribution, the full response including all SC fluctuations, and the contribution of BCS+Higgs fluctuations only. Diamonds denote the diamagnetic contribution, see Eq.\ \pref{kthzclean}, that is also present in the perfectly clean case ($V=0$), while circles denote the paramagnetic contribution, that only arises in the presence of disorder, see Eq.\ \pref{kthzdirty}. 
At the lowest disorder level $V/t=0.1$, the Higgs contribution (blue diamonds/circles) is quantitatively negligible, and for $V/t=0.5$ it only adds a correction at most of order of 30$\%$ of the BCS one, while phase modes 
(green diamonds/circles) give a sizeable contribution already at $V/t=0.1$. As a consequence, one can safely conclude that THG measurements in cuprates should be ascribed to the BCS response,  further enhanced by the contribution of phase fluctuations, while the Higgs response is largely subleading.  For what concerns the polarization dependence, one can see that the paramagnetic part becomes rapidly predominant at $\theta=0$, but the overall modulation of the two contributions has a similar strength, so that the overall response is pretty much isotropic at $V/t=0.1$, see Fig.\ \ref{polariz}, and only slightly modulated at $V/t=0.5$,  in excellent agreement with THG results, see Fig.\ \ref{polarization}b. 

Finally, we would like to comment on the polarization dependence expected in THz pump-optical probe measurements. The relevant processes for this kind of experiments are still depicted by the diagrams in Fig.\ \ref{diagrams}, but the frequencies running in the fermionic loops are different with respect to the case of simple THG experiments. Indeed, for an optical probe two external lines carry a large frequency in the visible, leading to a large frequency running in the internal loop. As observed in Ref.\ \cite{silaev_prb19}, in this situation disorder effects are expected to be less efficient in triggering a finite paramagnetic response, in contrast to what happens when only a frequency of the order of the THz pump field is involved. As a consequence, one can speculate that for THz Kerr effect measurements the relevant non-linear kernel can be a combination of the diamagnetic response plus only part of the paramagnetic one. If this is the case, it can be worth analyzing separately the angular dependence of the two contributions, as done in Fig.\ \ref{polariz} for the BCS part only. Notice that while for pump-probe experiments one can identify separately the $K_{B1g}$ and $K_{B2g}$ contributions by tuning independently the pump $\theta_P$ and the probe $\theta_S$ angles,  see Eq.\ \pref{eq:js}, in the case of THG experiments $j_\pll^{NL}$ depends only on the pump angle $\theta$, see Eq.\ \pref{eq:fit}, leaving some ambiguity in the identification of the various angle-dependent terms. In the specific case of our calculations we will fit the diamagnetic term with the sum of a $K_{A1g}$ and a $K_{B1g}$ contribution, by using the fact that for these Kubo-like diagrams the $K_{B2g}$ term in the clean limit can only scale with the sub-leading $t'$ next-nearest-neighbors hopping term. On the other hand, for paramagnetic-like diagrams one cannot establish a-priori a prevalence of one asymmetric channel over the other, so one can only fix in principle the relative weight of the combination of two terms, by rewriting e.g. Eq.\ \pref{eq:fit} as $j_{\parallel}^{NL}(\theta)=K_{A1g}+K_{B2g}+(K_{B1g}-K_{B2g})\cos^2(2\theta)
$, or the analogous expression where only a $\sin^2(2\theta)$ is left. Once clarified such an ambiguity, and in order to simplify the analysis, we decided to conventionally fit the paramagnetic term as the sum of a $K_{A1g}$ and $K_{B2g}$ terms only, by simply observing that the signal increases as $\theta$ increases. The relative weights of the various channels are reported in Table \ref{tab:table2} and \ref{tab:table3}. As one can see, the diamagnetic term has a sizeable $B_{1g}$ component with an increasing ratio $K_{B1g}/K_{A1g}$ as doping increases, especially for larger disorder. Even though these ratios are larger than the experimental findings of Ref.\ \cite{shimano_prl18}, one could expect that a partial compensation from the paramagnetic channel can explain the difference with THG measurements, and account for the observed doping dependence of the $K_{B1g}/K_{A1g}$ ratio. Finally, it is worth mentioning that a third possible mechanism has been recently proposed in Ref.\ \cite{gabriele_natcomm21}, based on two-plasmon excitation processes that are beyond the approximation studied here. By accounting for this additional channel within an $XY$ model description of plasma modes, one finds an additional contribution to the non-linear kernel having $K_{B1g}/K_{A1g}=0.5$. So far, the quantitative relevance of this effect with respect to the BCS response has not been estimated, making a direct comparison with experiments difficult. On the other hand, since the energy scale setting the strength of two-plasmon excitations is the superfluid stiffness, one would expect a larger contribution in overdoped samples, where it becomes quantitatively larger. A closer analysis of this problem, along with a direct estimate of the effective relevance of paramagnetic processes for the THz Kerr effect, will certainly help elucidating the nature of the THz non-linear response in cuprate superconductors, and will thus deserve future work. \\
\\

\section*{Tables}

\begin{table}[H]
\small
\begin{tabular*}{0.48\textwidth}{@{\extracolsep{\fill}}lllllll}
\hline
&
\multicolumn{3}{c}{$V/t=0.1$} &
\multicolumn{3}{c}{$V/t=0.5$}\\
\hline
& $K_{A1g}$ & $K_{B1g}$ & $K_{B1g}/K_{A1g}$ & $K_{A1g}$ & $K_{B1g}$ & $K_{B1g}/K_{A1g}$\\
\hline
$p=0.1$ & 0.038 & 0.135 & 3.545 & 0.054 & 0.086 & 1.576\\
$p=0.2$ & 0.040 & 0.150 & 3.732 & 0.035 & 0.096 & 2.756\\
\hline
\end{tabular*}
\caption{Results of the fitting procedure for the diamagnetic BCS-only contribution to $j_{\parallel}^{NL}(\theta)$ through Eq.\ \pref{eq:fit}, obtained by conventionally setting $K_{B2g}=0$.}
\label{tab:table1}
\end{table}

\begin{table}[H]
\small
\begin{tabular*}{0.48\textwidth}{@{\extracolsep{\fill}}lllllll}
\hline
&
\multicolumn{3}{c}{$V/t=0.1$} &
\multicolumn{3}{c}{$V/t=0.5$}\\
\hline
& $K_{A1g}$ & $K_{B2g}$ & $K_{B2g}/K_{A1g}$ & $K_{A1g}$ & $K_{B2g}$ & $K_{B2g}/K_{A1g}$\\
\hline
$p=0.1$ & 0.354 & 0.185 & 0.521 & 1.487 & 0.587 & 0.395\\
$p=0.2$ & 0.278 & 0.162 & 0.581 & 1.152 & 0.503 & 0.437\\
\hline
\end{tabular*}
\caption{Results of the fitting procedure for the paramagnetic BCS-only contribution to $j_{\parallel}^{NL}(\theta)$ through Eq.\ \pref{eq:fit}, obtained by conventionally setting $K_{B1g}=0$.}
\label{tab:table2}
\end{table}

\begin{table}[H]
\small
\begin{tabular*}{0.48\textwidth}{@{\extracolsep{\fill}}lllll}
\hline
&
\multicolumn{2}{c}{$V/t=0.1$} &
\multicolumn{2}{c}{$V/t=0.5$}\\  
\hline
& $K_{B1g}/K_{A1g}$ & $K_{B2g}/K_{A1g}$ & $K_{B1g}/K_{A1g}$ & $K_{B2g}/K_{A1g}$\\
\hline
$p=0.1$ & 0.344 & 0.470 & 0.055 & 0.381 \\
$p=0.2$ & 0.470 & 0.508 & 0.081 & 0.424 \\
\hline
\end{tabular*}
\caption{Results of the fitting procedure for the sum of diamagnetic and paramagnetic BCS-only contributions to $j_{\parallel}^3(\theta)$. The $A1g$ component is taken as the sum of the separate results from diamagnetic- and paramagnetic-only contributions.}
\label{tab:table3}
\end{table}



\section*{Acknowledgements}
This work has been supported by the Sapienza University
via Ateneo 2019 RM11916B56802AFE and Ateneo 2020 RM120172A8CC7CC7, by the Italian MIUR project PRIN 2017 No. 2017Z8TS5B. G.S. acknowledges financial support from the Deutsche Forschungsgemeinschaft under SE 806/19-1.

\bibliography{Literature.bib} 
\end{document}